\RequirePackage{lineno}

\documentclass[aps,prl,twocolumn,showpacs,superscriptaddress,groupedaddress,nofootinbib]{revtex4-1}  
\usepackage{graphicx}  
\usepackage{dcolumn}   
\usepackage{bm}        
\usepackage{amssymb}   
\usepackage{verbatim}
\usepackage{color}
\usepackage[english]{babel}
\usepackage{hyperref}
\usepackage{float}

\begin{document}



\title{Fabrication of 3D-printed PCTFE material cups for dynamic nuclear polarization target at cryogenic temperatures}

\author{K. J.~McGuire} \email[Corresponding author: ]{kjm1042@wildcats.unh.edu}\affiliation{University of New Hampshire, Durham, NH, 03824, USA} 
\author{E.~Long} \affiliation{University of New Hampshire, Durham, NH, 03824, USA} 
\date{\today}

\begin{abstract}
We present a novel method of 3D printing with the fluoroplastic Kel-F (PCTFE) that was used to create target material cups for dynamic nuclear polarization (DNP) experiments. Kel-F is used in DNP targets because it has several properties that make it well-suited to this purpose: transparency to millimeter-waves, plasticity at cryogenic temperatures, and the absence of hydrogen that would add an unwanted background to NMR signals used to measure proton polarization. A custom filament production device called the Filatizer was developed that processes commercially available rods of Kel-F into a 1.75-mm diameter filament that was 3D printed using a Prusa i3 Mk2.5S modified to achieve the high temperatures ($\sim 400^{\circ}$C) needed to melt the material and with copper-based components replaced to reduce material decomposition. This printing process does not significantly alter Kel-F's properties, and we demonstrate the first 3D-printed Kel-F target cups successfully used in DNP enhancement. 
\end{abstract}

\maketitle

Dynamic nuclear polarization (DNP) is a standard technique for creating a polarized target for nuclear physics experiments~\cite{Crabb:1997cy}. In DNP, the target is placed in a high magnetic field (typically $\sim 5$~T), cooled to cryogenic temperatures ($<1.5$~K), and exposed to microwave radiation to induce spin transfer between unpaired electrons in the sample and the target nucleons. Nuclear magnetic resonance (NMR) techniques are used to measure the resulting polarization of the nucleons.

DNP is performed at temperatures approaching 1~K to drive up the thermal equilibrium (TE) polarization of nuclei. For spin-1/2 particles, the TE polarization is described by Maxwell-Boltzmann statistics as
\begin{equation}
P_{TE}(1/2) = \tanh \left( \frac{\hbar \gamma B}{2k_{B} T} \right),
\end{equation} 
where $\gamma$ is the gyromagnetic ratio of a spin-1/2 particle, $B$ is the magnitude of the applied magnetic field, $T$ is the temperature of the system, and $k_{B}$ is the Boltzmann constant. For a system in thermal equilibrium at 1 K in a 5 T field, $P_{e}= 99.8\%$ for electrons and $P_{p}=0.51\%$ for protons. The large magnitude difference between the electron and proton polarization is due to the proton's much smaller magnetic moment. 

In DNP, electron polarization is transferred to the target nucleons such that, in an ideal case, the maximum polarization of the target approaches that of $P_{e}$ in thermal equilibrium. The target is irradiated at a frequency corresponding to the sum of the spin resonance frequencies of the electron and target nucleon in the magnetic field,
\begin{equation}
\nu = B/2\pi\left(\gamma_{e} + \gamma_{N}\right).
\end{equation} 
In the case of protons in a 5 T field, $\nu$ $\approx$ 140 GHz, which corresponds to a wavelength of $\sim$2.14 mm (``mm-waves").

\begin{figure}
  \includegraphics[width=1.0\linewidth]{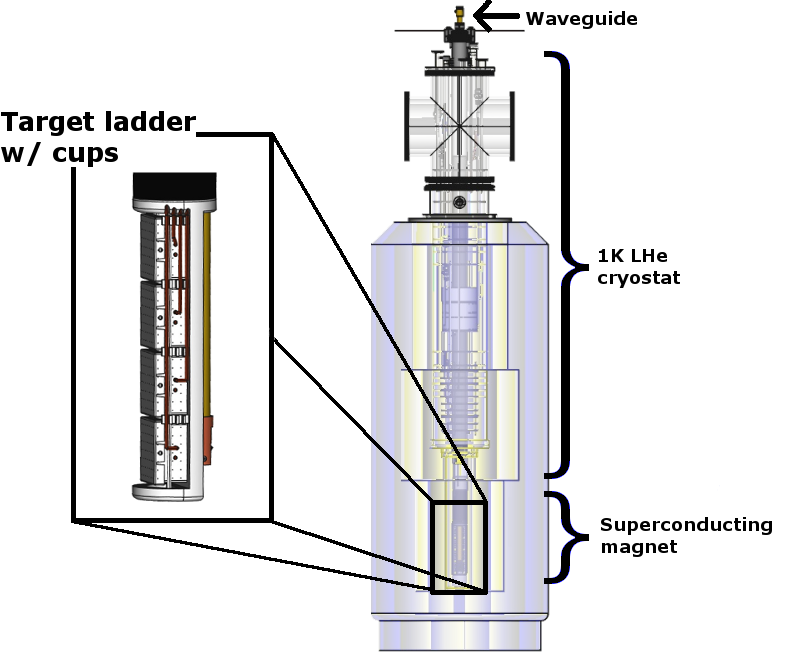}
  \caption{Dynamic nuclear polarization system. A target ladder with material cups, thermometry, and NMR circuit is placed at the center of a superconducting electromagnet in a cryostat. DNP is driven via microwaves that are transmitted along the waveguide and irradiate the target material.} 
  \label{fig:DNPsystem}
\end{figure}

The target material is housed in small cups affixed to a target ``ladder" and submerged in a superfluid helium cryostat (Fig.~\ref{fig:DNPsystem}). Because of the extreme conditions under which DNP is performed, a suitable material must be chosen for the target cups.  

The fluoroplastic Kel-F (polychlorotrifluoroethylene [PCTFE]) has several properties that make it a favorable material for the DNP target cups: 1) it retains its plasticity at $<1$~K temperatures; 2) it is highly transparent to mm-wave radiation; and 3) it contains no hydrogen that would introduce a background into the NMR signal used to measure polarization. For these reasons, Kel-F has traditionally been used in DNP experiments~\cite{Pierce:2013pua}; however, the Kel-F components had to be machined using subtractive manufacturing techniques. 

Additive manufacturing (i.e., 3D printing) offers several advantages over machining, including rapid prototyping and the ability to manufacture parts with complex geometries and fine details that would be difficult or impossible to machine. Using custom methods developed in-house, the University of New Hampshire's (UNH) Nuclear Physics Group has successfully printed Kel-F target cups and demonstrated their favorable performance in a DNP target system.

Kel-F is a thermally unstable fluoroplastic and special care must be taken when heat-processing the material as it will begin to degrade and release chlorine and/or fluorine gases upon reaching its melting point of 212$^{\circ}\mathrm{C}$~\cite{PCTFE1}. 
To minimize the risk of contamination, all fluoroplastic processing was performed in a fume hood. Care was taken to ensure that all heated brass or copper components were replaced with stainless steel, aluminum, or nickel-coated components, as copper acts as a catalyst for the decomposition into hydrofluoric acid (HF) and hydrochloric acid (HCl) \cite{decomp}. Prior to filament production, the University of New Hampshire's Office of Environmental Health and Safety performed air and surface contamination tests that were analyzed by Absolute Resource Associates. During these tests, Kel-F was heated to various temperatures up to 400$^{\circ}$C and held at $>300^{\circ}$C for approximately 30 minutes to maximize decomposition. These tests yielded a maximum of 0.1 mg/L fluoride and 0.5 mg/L chloride, which fall below the U.S. Environmental Protection Agency limits of 4 mg/L for fluoride and 250 mg/L for chloride recommended in drinking water~\cite{EPA}. 





A major challenge of 3D printing with Kel-F is finding the optimal temperature and speed at which to melt and extrude the material. Kel-F is a semi-crystalline polymer, and its degree of crystallinity can vary considerably and increases with prolonged heat exposure~\cite{PCTFE1}. The higher the crystallinity, the more glassy and brittle Kel-F becomes, so it is necessary to melt, extrude, and cool the material as quickly as possible to minimize crystallization. Because Kel-F has a high melt viscosity~\cite{PCTFE1}, it must be processed at a temperature well above its melting point. However, as noted above, Kel-F is also thermally unstable and will decompose rapidly at high temperatures. Hence there is a small temperature and time window in which Kel-F can be processed without significantly altering its desired properties.

\begin{figure}
  \includegraphics[width=0.75\linewidth]{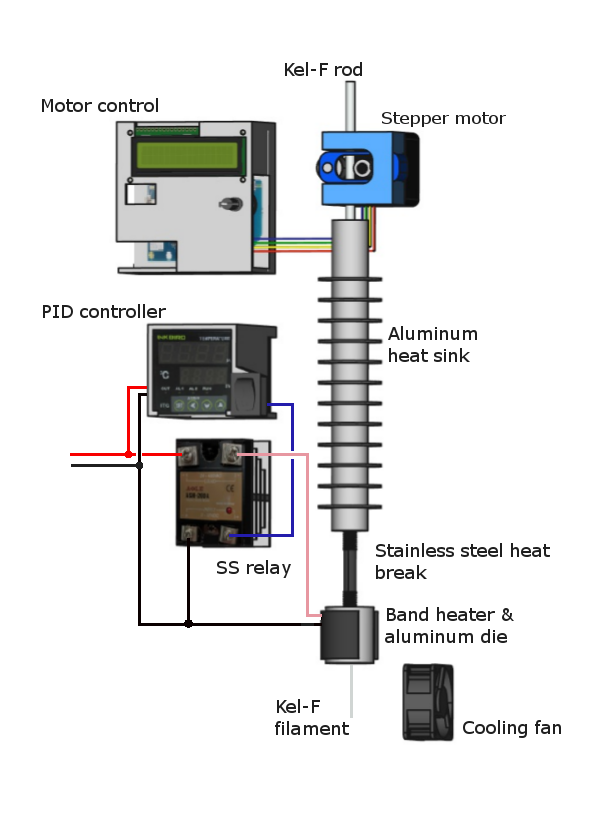}
  \caption{``Filatizer" filament production system. It uses a stepper motor to feed the bulk material through a heat-sink and into a hot end, where it is extruded through an aluminum die at high temperature. An external fan cools the extruded filament to reduce crystallization.} 
  \label{fig:filatizer}
\end{figure}

A new device that processes bulk 0.187-inch diameter Kel-F rods into printable 1.75-mm diameter filament was developed and dubbed the ``Filatizer." An overview of the system is shown in Fig.~\ref{fig:filatizer}. It consists of a stepper motor and controller that feeds the bulk rod into an aluminum heat sink to keep the material not currently being processed at a low temperature, a stainless steel heat break, and a heated aluminum die with an external diameter of 1 inch and an internal conical hole that reduces in size from 0.2 inches to $\sim$2 mm in diameter. The die was heated using a 150~W band heater. The temperature was monitored and controlled via a thermocouple and PID controller that drives a solid-state relay that powers the heater band. After extrusion, the filament was cooled using a small, power-adjustable 12~V fan and wound onto a spool using the commercially available Filawinder~\cite{Filastruder}. The diameter of the filament was determined by the amount of cooling at the die and the spooling rate, which was controlled using a laser positioning system that is standard with the Filawinder.

\begin{figure}
\includegraphics[width=\linewidth]{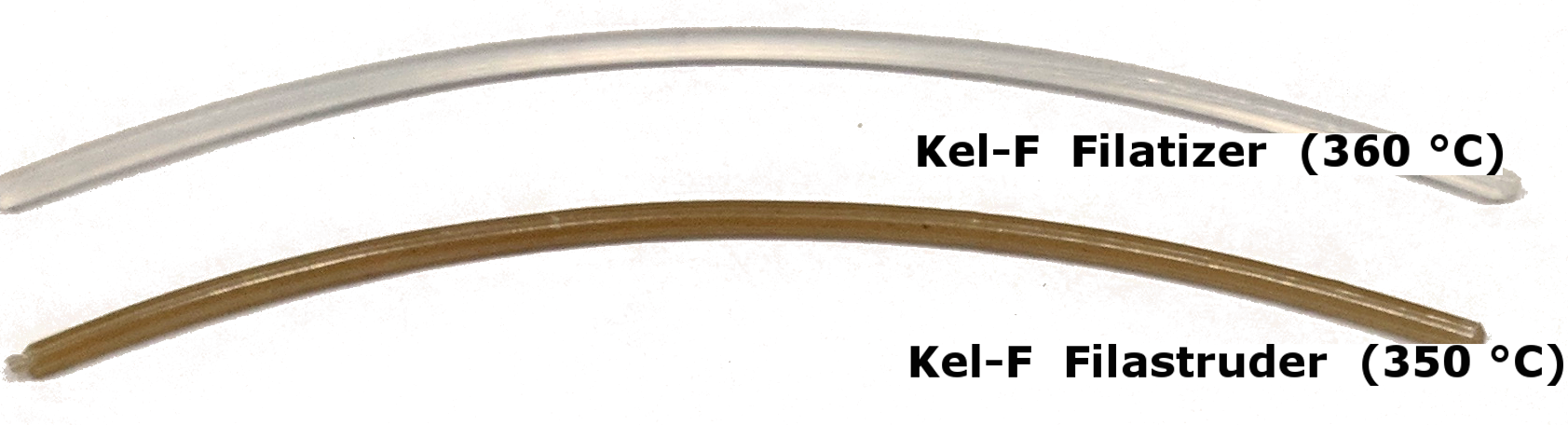}
\caption{Kel-F filament produced using the Filatizer system described in this paper largely maintains its original properties (top) compared to filament produced with the commercially available Filastruder that resulted in obvious material decomposition and discoloration (bottom).}
\label{fig:filCompare}
\end{figure}

The Filatizer significantly reduced the volume of melted material in the hot end compared with commercially available filament-making devices such as the Filastruder~\cite{Filastruder}, which increased the processing speed of the material and resulted in cleaner filament with less crystallization as shown in Fig.~\ref{fig:filCompare}.

Because of Kel-F's high melt viscosity, the speed and temperature at which high-quality filament can be extruded is largely dependent on the diameter of the die. Good results were obtained at several different settings given in Tab.~\ref{tab:filatizer}.

\begin{table}
\caption{Settings for Kel-F filament production using the Filatizer.}
\begin{ruledtabular}
\begin{center}
\begin{tabular}{ c  c  c  c } 
	Die  & Temp.  & Speed  & Filament \\ 
	(mm) & ($^{\circ}$C) & (cm/min) & (mm) \\ \hline
	2.00 & 360 & 13 & 1.75$\pm0.1$  \\ 
	1.65 & 380& 17 & 1.7$\pm0.1$  
\end{tabular}
\label{tab:filatizer}
\end{center}
\end{ruledtabular}
\end{table}

Kel-F printing was performed using a modified Prusa i3 MK2.5S printer and the included Slic3r software to generate gcode files sent to the printer~\cite{Prusa}. To accommodate temperatures above 300$^{\circ}\mathrm{C}$, the aluminum heater block was replaced with nickel-coated copper~\cite{e3d}, and the printer was modified to use a Type K thermocouple in place of the thermistor at the hot end. To prevent catalytic decomposition, the default brass extrusion nozzle was replaced with hardened steel.

Kel-F, like other fluoroplastics, is a slippery material with a low coefficient of static friction~\cite{PCTFE2}. One of the major challenges of printing with the material was getting the component to adhere to the printer bed and remain affixed for the duration of the print. Kel-F will not adhere directly to the printer bed, at least not at bed temperatures below 120$^{\circ}\mathrm{C}$, which is the upper limit of the Prusa printer. The material might adhere better at a bed temperature closer to the melting point of Kel-F (212$^{\circ}\mathrm{C}$); however, such high temperatures would dramatically slow the cooling of the print and would likely result in the material losing its plasticity as it cooled. Instead, Kel-F was printed atop a layer of high-temperature silicon glue on adhesive transfer tape~\cite{McMaster}. This adhesive is easily applied to the print bed but leaves a residue on the bottom of the print that must be cleaned in post-processing.

\begin{figure}
\centering
\includegraphics[width=8cm]{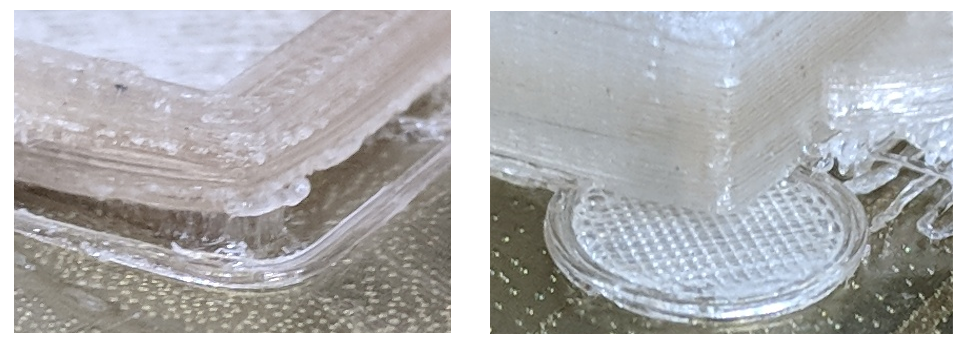}
\caption[Kel-F bed adhesion]{Kel-F printer bed adhesion was improved by adding circular ``feet" to sharp edges. }
\label{fig:feet}
\end{figure}

The problem of bed adhesion is compounded by the tendency of the material to curl up and off of the printer bed as the material cools. The effect is most pronounced at sharp corners and was mitigated by adding thin circular ``feet" at the corners as shown in Fig.~\ref{fig:feet}, which increased the surface area at the print bed. The feet were trimmed from the component once the print was complete. This method worked better than printing the model with a brim or raft,
which did not prevent curling.     

\begin{figure}
  \includegraphics[width=1\linewidth]{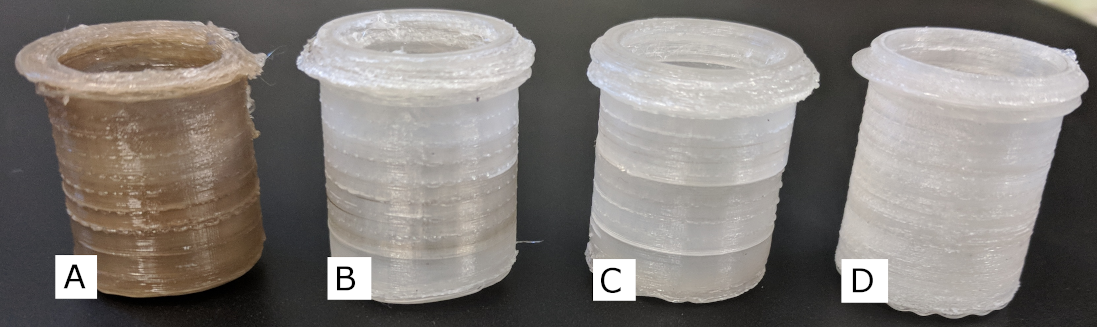}
  \caption[3D-printed Kel-F iterations]{3D-printed Kel-F DNP target cups. (A) was printed with filament from the Filastruder; (B) and (C) were printed with filament from the Filatizer and show considerable color improvement but contain regions that are glassy and brittle (visible as distinct lines); (D) was printed at a faster extrusion speed and the material retained its plasticity.} 
  \label{fig:iterations}
\end{figure}

Fine-tuning the print settings was imperative for a successful Kel-F print. While true of fused filament fabrication in general, this is especially true when printing with Kel-F, as even slight variations in print speed and temperature can dramatically affect the quality of a print. This is demonstrated in Fig.~\ref{fig:iterations}, which shows printed cups of various temperatures and speeds that resulted in different levels of material decomposition and crystallization.  Several print settings that minimized these negative effects and resulted in successful prints are given in Tab.~\ref{tab:printsettings}.  

\begin{table}
\caption{Settings for Kel-F printing on a modified Prusa i3 MK2.5S printer using Slic3r software to generate gcode.}
\begin{ruledtabular}
\begin{tabular}{  l  c  c }  
 Filament (mm)& 1.75$\pm0.1$ &  1.7$\pm0.1$\\ 
 
 Fill density ($\%$)& 100 &100  \\ 
 Flow rate ($\%$) & 80 & 70-80  \\ 
 Layer height (mm)& 0.2& 0.1\\ \hline
	 First layer: & & \\ 
	 
          \hspace{0.5cm} Temp. ($^{\circ}$C)&  & \\ 
\hspace{1cm}    Nozzle& 360 & 370\\ 
\hspace{1cm}    Bed& 115 & unheated  \\ 
\hspace{0.5cm}          Speed (mm/sec) & 6 & 7.5\\ 
\hspace{0.5cm}  Extrusion width (mm) & 0.42 & 0.42\\ \hline

          Subsequent layers:& & \\ 
          \hspace{0.5cm} Temp. ($^{\circ}$C)& &\\  
\hspace{1cm}    Nozzle& 360 &370\\ 
\hspace{1cm}    Bed& 115 &unheated\\           
          \hspace{0.5cm}    
          Speed (mm/sec) &&\ \\ 
\hspace{1cm}          Perimeter& 7.5 &15\\ 
\hspace{1cm}   Small perimeters       & 3 &6 \\ 
\hspace{1cm}   External perimeters    & 6  &12 \\  
\hspace{1cm}   Infill         & 9 & 18\\ 
\hspace{1cm}  Solid infill          & 7.5  &15  \\ 
\hspace{1cm}    Top solid infill       & 6   &12 \\ 
\hspace{1cm}     Gap fill       & 4.5 &9\\ 
\hspace{0.5cm} Extrusion width & 0.45 & 0.45\\ \hline
Print fan & Off & Off  \\ 
Notes:  &  \multicolumn{2}{l}{Printed atop silicon} \\
  &  \multicolumn{2}{l}{adhesive transfer tape} \\ 
    \end{tabular}
    \label{tab:printsettings}
\end{ruledtabular}
\end{table}

\begin{figure}
\centering
\includegraphics[width=0.85\linewidth]{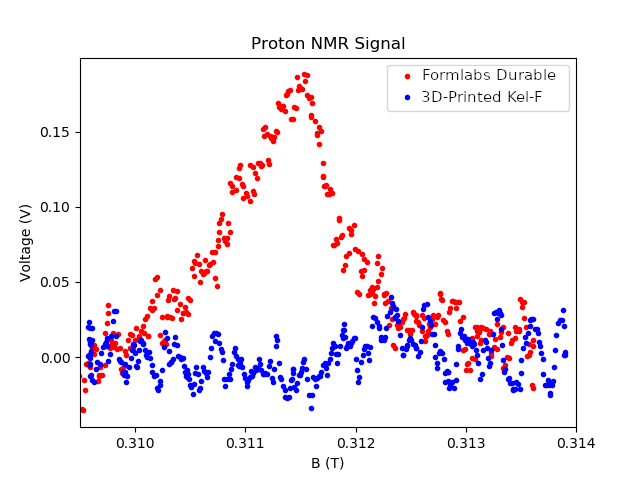}
\caption{Proton NMR analysis of a sample of heavily processed Kel-F showed no indication of hydrogen content. A similar measurement on Formlabs Durable resin yielded a significant proton NMR signal.}
\label{fig:rtnmr}
\end{figure}

To ensure that the 3D printing process did not add hydrogen to the printed components, a room temperature NMR analysis at $\sim13.2$~MHz was performed on a printed slug of Kel-F using a Teltron U188031 ESR/NMR system with a Teltron U189021 NMR probe. As shown in Fig.~\ref{fig:rtnmr}, this resulted in no observable proton signal for Kel-F. For comparison, an identical slug was printed using Formlab's Durable material~\cite{Formlabs}, which shows a clear proton signal. 

A prototype target ladder with Kel-F cups was printed and impact tested for robustness at cryogenic temperatures. This ladder was temperature cycled from 77~K to room temperature six times using liquid nitrogen and showed no noticeable difference in the Kel-F components. The same ladder was then brought to 77~K and immediately dropped from increasing heights. One cup that had a small region of crystalline Kel-F broke along that region when dropped from a height of 8~inches. Two cups detached from the ladder when dropped from 35~inches as the adhesive that held the Kel-F cups to the ladder failed, although the Kel-F cups remained intact. When dropped from a height of 4~feet the prototype target ladder shattered while the remaining Kel-F cups showed no signs of damage.

\begin{figure}[H]
  \includegraphics[width=1\linewidth]{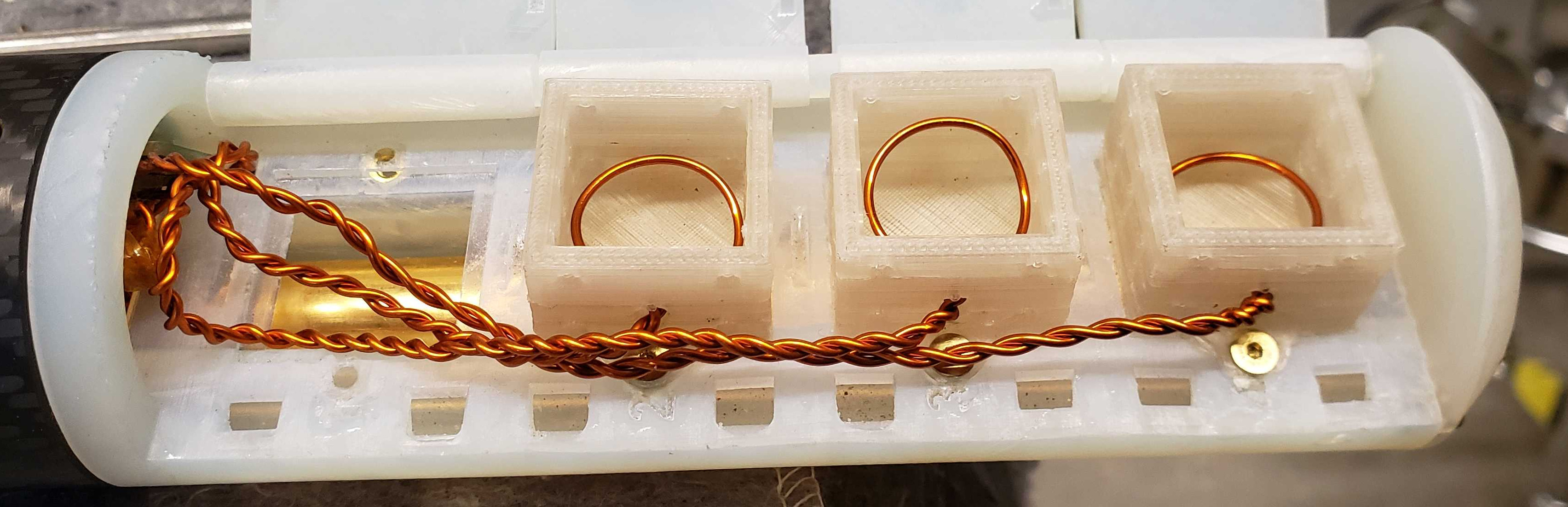}
  \caption{DNP target ladder with NMR coils and square 3D-printed Kel-F cups after cooling to 1~K and used for DNP enhancement. The cup on the far left was removed to allow access to the NMR circuit.} 
  \label{fig:ladder}
\end{figure}

Following these tests, square Kel-F target material cups were printed and installed on the UNH DNP target ladder as shown in Fig.~\ref{fig:ladder}. This ladder survived multiple 1~K temperature cycling and was successfully used within the UNH DNP system to hyperpolarize TEMPO-doped Araldite and butanol. The 140~GHz mm-waves were transmitted through the bottom of the printed Kel-F cups to drive polarization enhancement. Polarization enhancement results will be in a forthcoming publication.

In conclusion, we have created a procedure to 3D print Kel-F components. A new filament production system was presented that allows for fluoroplastic printing using a customized Prusa i3 Mk2.5S 3D printer. Components printed with this system have been successfully used as target material cups in a DNP system and survived multiple 1~K temperature cycling.

This work was supported by the U.S. Department of Energy under grant DE-FG02-88ER40410.


\bibliographystyle{apsrev4-1}
\bibliography{Fabrication_of_3D-printed_PCTFE_material_cups.bib}			

\end{document}